\documentclass[epj]{webofc}
\usepackage[varg]{txfonts}
\usepackage[english]{babel}
\usepackage[T1]{fontenc}
\usepackage{graphicx,color,subfigure}
\usepackage{amsmath,bm,amssymb}  
\usepackage{indentfirst}
\usepackage{multirow}
\usepackage[colorlinks=true,linkcolor=blue,citecolor=blue,urlcolor=blue]{hyperref}

\woctitle{21st International Conference on Few-Body Problems in Physics}

\begin{document}

\title{Strangeness in nuclei and neutron stars: a challenging puzzle}

\author{Diego Lonardoni\inst{1}\and
        Alessandro Lovato\inst{1} \and
        Stefano Gandolfi\inst{2} \and
        Francesco Pederiva\inst{3,4}
}

\institute{Physics Division, Argonne National Laboratory, Lemont, Illinois 60439, USA
\and
           Theoretical Division, Los Alamos National Laboratory, Los Alamos, New Mexico 87545, USA
\and
           Department of Physics, University of Trento, Via Sommarive 14, I-38123 Trento, Italy
\and
           INFN-Trento Institute for Fundamental Physics and Application, Trento, Italy
          }

\abstract{
The prediction of neutron stars properties is strictly connected to the employed nuclear interactions. The appearance of hyperons in the inner core of the star is strongly dependent on the details of the underlying hypernuclear force. We summarize our recent quantum Monte Carlo results on the development of realistic two- and three-body hyperon-nucleon interactions based on the available experimental data for light- and medium-heavy hypernuclei.
}

\maketitle

\section{Introduction}
Neutron stars are among the most compact and dense objects in the Universe, with typical masses up to $\sim2~M_\odot$ and radii $R\sim10$~km. Their central densities can be several times larger than nuclear saturation density, $\rho_0=0.16~\text{fm}^{-3}$. At densities larger than $\sim2\rho_0$, when the nucleon chemical potential is large enough, the conversion of nucleons into hyperons might become energetically favorable, and Pauli blocking would prevent hyperons from decaying by limiting the phase space available to nucleons. This would lead to a reduction of the Fermi pressure exerted by the baryons and to a softening of the equation of state (EOS). As a consequence, the maximum mass determined by the equilibrium condition between gravitational and nuclear forces would be reduced.

Currently there is no general agreement (even qualitative) among the predicted results for the EOS of strange baryonic matter and the maximum mass of a neutron star (NS), that violates the constraints given by the recently measured value of $\sim2~M_\odot$ of the millisecond pulsars PSR~J1614-2230~\cite{Demorest:2010} and PSR~J0348+0432~\cite{Antoniadis:2013}, see Ref.~\cite{Lonardoni:2015}. This apparent inconsistency between NS mass observations and theoretical calculations is a long standing problem known as the \emph{hyperon puzzle}. It has to be ascribed to a combination of an incomplete knowledge of the forces governing the system and to the concurrent use of approximate theoretical many-body techniques. Its solution requires a more accurate theoretical analysis and a thorough experimental investigation of the hyperon-nucleon force in a variety of systems, ranging from light to medium and heavy hypernuclei.

We employ a quantum Monte Carlo computational technique to tackle the general problem of the hyperon-nucleon interaction. The algorithm, auxiliary field diffusion Monte Carlo (AFDMC), has been introduced by Schmidt and Fantoni~\cite{Schmidt:1999} as an extension of the usual diffusion Monte Carlo method to deal in an efficient way with spin/isospin-dependent Hamiltonians. It has been successfully applied to solve the many-body problem in a non perturbative fashion for strongly correlated systems, both in the standard~\cite{Gandolfi:2011,Gandolfi:2012,Gandolfi:2014,Carlson:2015} and strange nuclear sector~\cite{Lonardoni:2013,Lonardoni:2014,Lonardoni:2015,Pederiva:2015}.

\section{Hypernuclei: from few- to many-body systems}
\begin{figure}[b]
\centering
\includegraphics[width=0.8\textwidth]{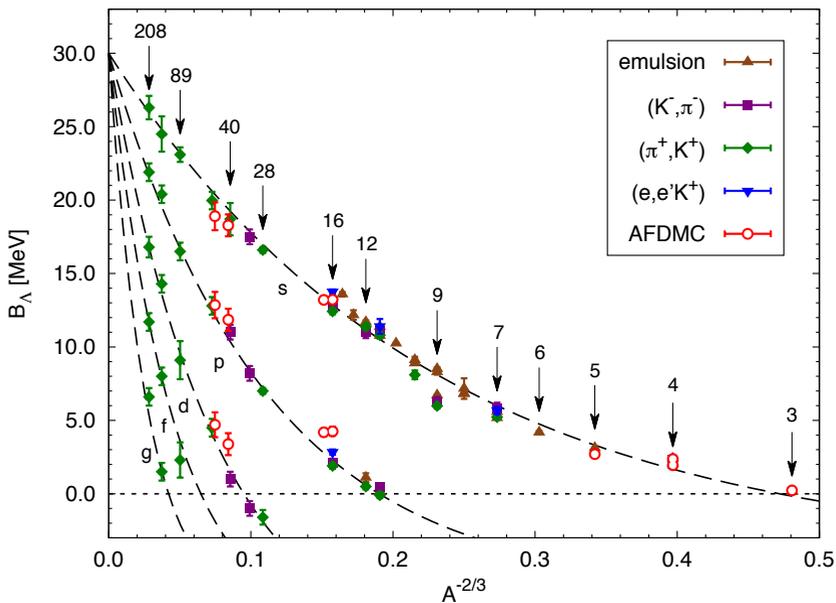}
\caption{Measured (see Ref.~\cite{Pederiva:2015} for the complete list of references) and computed $\Lambda$~separation energies as a function of $A^{-2/3}$. 
Results for the $\Lambda$~particle in different single particle states are shown.}
\label{fig1}
\end{figure}

As reported in Refs.~\cite{Lonardoni:2013,Lonardoni:2014,Pederiva:2015}, an accurate analysis of the $\Lambda$~separation energy $B_\Lambda$ of light- and medium-heavy hypernuclei has been recently carried out using a phenomenological hypernuclear interaction analog to the Argonne-Illinois nucleon-nucleon force~\cite{Bodmer:1984,Bodmer:1985,Bodmer:1988,Usmani:1995,Usmani:1995_3B,Usmani:1999,Usmani:2008,Imran:2014}. The $\Lambda N$ two-body component of this interaction has been fit to the existing $\Lambda p$ scattering data. The $\Lambda NN$ three-body sector instead has been re-fit to the available $\Lambda$~separation energies of single-$\Lambda$~hypernuclei over a wide mass range by means of AFDMC calculations. It has been shown that the repulsive nature of the three-body hyperon-nucleon interaction is the key to satisfactorily reproduce the ground state properties of hypernuclei from light- to medium-heavy within a unique theoretical framework. Fig.~\ref{fig1} shows the AFDMC results for $B_\Lambda$ when using the complete $\Lambda N+\Lambda NN$ force. The agreement with experimental data is good over all the mass range investigated and for the $\Lambda$~particle in different single particle states. For more details see Refs.~\cite{Lonardoni:2013,Lonardoni:2014,Pederiva:2015}.

In the light sector, a two-body $\Lambda N$ charge symmetry breaking interaction has been also employed~\cite{Bodmer:1985,Lonardoni:2014}. As shown in Tab.~\ref{tab1}, AFDMC results reproduce the hyperon separation energies and energy splitting of the mirror hypernuclei $^4_\Lambda$H and $^4_\Lambda$He, according to the old emulsion data~\cite{Juric:1973}. $^3_\Lambda$H is weakly bound compared to deuterium and the neutral hypernucleus $^3_\Lambda$n is predicted to be unbound, in agreement with other theoretical results~\cite{Hiyama:2014,Gal:2014,Richard:2015}, contradicting the possible experimental evidence reported in Ref.~\cite{Rappold:2013}.

\renewcommand{\arraystretch}{1.6}
\begin{table}[h]
\centering
\caption{AFDMC and experimental~\cite{Juric:1973} $\Lambda$~separation energies for the lightest hypernuclei. In the case of $A=4$ the separation energy difference $\Delta B_\Lambda$ is also shown. Evidence for the neutral hypernucleus $^3_\Lambda$n has been reported in Ref.~\cite{Rappold:2013}, but no $B_\Lambda$ value has been reported. All energies are in MeV.}
\label{tab1}
\begin{tabular}{l|cc|cc}
System & $B_\Lambda$ & $B_\Lambda^{exp}$ & $\Delta B_\Lambda$ & $\Delta B_\Lambda^{exp}$ \\
\hline\hline
$^3_\Lambda$n $\left(\displaystyle\tfrac{1}{2}^+\right)$ & \hspace{0.25em}unbound\hspace{0.25em} & unknown & \multirow{2}{*}{---} & \multirow{2}{*}{---} \\   
$^3_\Lambda$H $\left(\displaystyle\tfrac{1}{2}^+\right)$ & \hspace{0.25em}0.23(9)\hspace{0.25em} & 0.13(5) & & \\
\hline
$^4_\Lambda$H $\left(\displaystyle 0^+\right)$ & 1.95(9) & 2.04(4) & \multirow{2}{*}{0.42(11)} & \multirow{2}{*}{0.35(6)} \\ 
$^4_\Lambda$He $\left(\displaystyle 0^+\right)$& 2.37(9) & 2.39(3) & & \\
\end{tabular}
\end{table}

In order to validate the phenomenological approach adopted in quantum Monte Carlo calculations for hypernuclei, a benchmark among different few-body methods has been recently proposed. In tab.~\ref{tab2} we report preliminary results for $^5_\Lambda$He for two different exact methods, AFDMC and Nonsymmetrized Hyperspherical Harmonics (HSHH)~\cite{Deflorian:2013}. In both strange and non-strange sectors the employed potentials are at two-body level only. Results agree among different methods whether using a soft (Minnesota~\cite{Thompson:1977}) or hard (Argonne V4$^\prime$~\cite{Wiringa:2002}) $NN$ potential.

\renewcommand{\arraystretch}{1.6}
\begin{table}[h]
\centering
\caption{Binding energies and $\Lambda$~separation energies for $^5_\Lambda$He for different combinations of nucleon-nucleon and hyperon-nucleon potentials and different methods. All energies are in MeV.}
\label{tab2}
\begin{tabular}{c|lc|lc}
$^{5}_\Lambda$He & \multicolumn{2}{c}{AFDMC} & \multicolumn{2}{|c}{NSHH} \\
\hline\hline  
\multirow{2}{*}{$NN$ Minn~\cite{Thompson:1977} $+\,\Lambda N$} & $E$         & -37.69(8) & $E$         & -37.77(10) \\
                                          & $B_\Lambda$ & 6.95(9)   & $B_\Lambda$ & 6.99(10) \\ 
\hline
\multirow{2}{*}{$NN$ AV4$^\prime$~\cite{Wiringa:2002} $+\,\Lambda N$} & $E$         & -39.46(12) & $E$         & -39.54(10) \\
                                                  & $B_\Lambda$ & 6.70(16)   & $B_\Lambda$ & 6.84(10) \\ 
\end{tabular}
\end{table}

\section{Neutron stars: from hypernuclei to the infinite medium}
Precise AFDMC calculations have been performed to first assess the parameters of the interaction from measured hyperon separation energies, and then to extrapolate the results to the case of neutron matter with strangeness. However, parametrizations of the potential predicting relatively small differences in the $\Lambda$~separation energies of hypernuclei give very different results for the properties of the infinite medium~\cite{Lonardoni:2015}. The resulting EOS spans the whole regime extending from the appearance of a substantial fraction of hyperons at $\sim2\rho_0$ to the absence of $\Lambda$ particles in the entire density range of the star. This yields a sizable effect on the predicted NS structure. Therefore, the derivation of realistic hypernuclear potential models is of paramount importance to properly assess the role of hyperons in NSs and reconcile theoretical predictions with astrophysical observations. This demands a precise and systematic experimental investigation of hypernuclear properties over a wide range of masses. For instance, a recent study of the isospin dependence of the present three-body hyperon-nucleon force shows the difficulty in extracting the information on the Hamiltonian from currently available experimental information on hypernuclei~\cite{Pederiva:2015}.

\begin{acknowledgement}
This work was supported by the U.S. Department of Energy, Office of Science, Office of Nuclear Physics, under the NUCLEI SciDAC grant (D.L., A.L., S.G.), by the Department of Energy, Office of Science, Office of Nuclear Physics, under Contract No. DE-AC02-06CH11357 (A.L.), and by DOE under Contract No. DE-AC52-06NA25396 and Los Alamos LDRD grant (S.G.). This research used resources of the National Energy Research Scientific Computing Center (NERSC), which is supported by the Office of Science of the U.S. Department of Energy under Contract No. DE-AC02-05CH11231, and computing time provided by Institutional Computing (IC) at LANL.
\end{acknowledgement}

\end{document}